\documentclass[conference, 9pt]{IEEEtran}

\usepackage{cite}
\usepackage{graphicx}

\usepackage{amsmath}
\usepackage{array}
\usepackage{booktabs}
\usepackage{caption}
\usepackage{subcaption}
\usepackage{floatrow}
\usepackage{enumitem}
\setlist{nolistsep}
\begin{document}
\bstctlcite{IEEEexample:BSTcontrol}
%
%
\title{On the Intrinsic Robustness of NVM Crossbars Against Adversarial Attacks}

\author{
\IEEEauthorblockN{Deboleena Roy, Indranil Chakraborty, Timur Ibrayev, Kaushik Roy}
\IEEEauthorblockA{Department of Electrical and Computer Engineering\\
Purdue University, West Lafayette, Indiana, USA\\
\{roy77, ichakra, tibrayev, kaushik\}@purdue.edu}
}
\maketitle

\begin{abstract}
The increasing computational demand of Deep Learning has propelled research in special-purpose inference accelerators based on emerging non-volatile memory (NVM) technologies. Such NVM crossbars promise fast and energy-efficient in-situ Matrix Vector Multiplication (MVM) thus alleviating the long-standing von Neuman bottleneck in today's digital hardware. However, the analog nature of computing in these crossbars is inherently approximate and results in deviations from ideal output values, which reduces the overall performance of Deep Neural Networks (DNNs) under normal circumstances. In this paper, we study the impact of these non-idealities under adversarial circumstances. We show that the non-ideal behavior of analog computing lowers the effectiveness of adversarial attacks, in both Black-Box and White-Box attack scenarios. In a non-adaptive attack, where the attacker is unaware of the analog hardware, we observe that analog computing offers a varying degree of intrinsic robustness, with a peak adversarial accuracy improvement of 35.34\%, 22.69\%, and 9.90\% for white box PGD ($\epsilon$=1/255, iter=30) for CIFAR-10, CIFAR-100, and ImageNet respectively. We also demonstrate ``Hardware-in-Loop" adaptive attacks that circumvent this robustness by utilizing the knowledge of the NVM model. \end{abstract}

\section{Introduction}

Deep Learning \cite{lecun2015deep} has emerged as a popular, versatile machine learning methodology that can be applied to a wide range of optimization tasks, such as computer vision \cite{voulodimos2018deep}, natural language processing \cite{young2018recent}, recommender systems \cite{cheng2016wide}, etc. As our reliance on deep learning increases, so does our need to build secure, reliable, efficient frameworks for executing its intensive computational requirements.

To accommodate the growing computational needs of Deep Neural Networks (DNNs) special-purpose accelerators such as GoogleTPU \cite{jouppi2017datacenterold}, Microsoft BrainWave \cite{chung2018serving}, and NVIDIA V100 \cite{nvidiav100} have been proposed. These systems perform efficient Matrix-Vector Multiplication (MVM) operations, the key computational kernel in DNNs, by co-locating memory and processing elements. Despite their success, the saturating scaling trends of digital CMOS \cite{xu2018scaling} has garnered interest in Non-Volatile Memory (NVM) technologies such as RRAM \cite{wong2012metal}, PCRAM \cite{wong2010phase} and Spintronics \cite{fong2015spin}. The memory element in these technologies is arranged in a crossbar fashion to enable efficient MVM computations in the analog domain inside the memory array. Such an in-memory computing primitive can significantly lower power and latency compared to digital CMOS \cite{chakraborty2020resistive}. Promises offered by the NVM crossbars have propelled significant research in designing analog computing based accelerators, such as PUMA \cite{shafiee2016isaac,ankit2019puma}.
In an analog computing hardware, the output of an MVM operation is sensed as a summation of currents through resistive NVM devices arranged in a crossbar, and hence are prone to errors due to non-ideal behavior of the crossbar and its peripheral circuits. Such errors are hard to model due to the interdependence of multiple analog variables (voltages, currents, resistances) in the crossbar. These deviations result in overall performance degradation of the DNN implementation \cite{chakraborty2020geniex}. Several works have explored various techniques to counteract the impact of these non-idealities \cite{liu2017rescuing, chakraborty2018technology}. 

On the flip side, even though the changes in DNN activations arising from non-idealities is difficult to model, it can potentially lead to adversarially robust DNN implementations. Adversarial images are generated by estimating the gradients of the DNN with respect to its input, and carefully perturbing the images in the direction of maximum change in the classifier output \cite{szegedy2013intriguing, goodfellow2014explaining}. To counter such attacks, several techniques that rely on gradient obfuscation have been previously proposed \cite{dhillon2018stochastic, buckman2018thermometer, athalye2018obfuscated}. In this work, we explore how non-ideal NVM crossbars have a similar intrinsic effect of gradient obfuscation. We implement DNNs on the PUMA architecture, which is composed of thousands of NVM crossbar based MVM units (MVMUs). The aforementioned errors occur at the output of these internal MVMUs, which are practically inaccessible to a third party user, such as the software designer or even an attacker. Moreover, the nature of the errors depends heavily on the technology,
which might not be fully disclosed by the manufacturer. We study two distinct scenarios, one where the attacker does not have access to custom NVM hardware and generates attacks based on "accurate" digital hardware, and the other where the attacker generates attacks with the NVM hardware in loop.
The main contributions of this work are as follows:
\begin{itemize}
    \item We demonstrate that adversarial attacks crafted without the knowledge of the hardware implementation are less effective in both black box and white box scenarios.
    \item We tested multiple variants of NVM crossbars, and show that the degree of intrinsic robustness offered by the analog hardware is in proportion to its degree of non-ideal behavior. 
    \item We show that ``Hardware-in-Loop'' adaptive adversarial attacks are more effective, as the attacker can now account for the non-ideal computations when crafting the adversarial examples. The degree of success depends on what hardware is available to the attacker and how similar it is to the target DNN's hardware implementation.
\end{itemize}

\section{Background and Related Work}

\subsection{In-memory Analog Computing Hardware}

\begin{figure}[h!]
	\centering
	\includegraphics[width=0.8\textwidth, keepaspectratio]{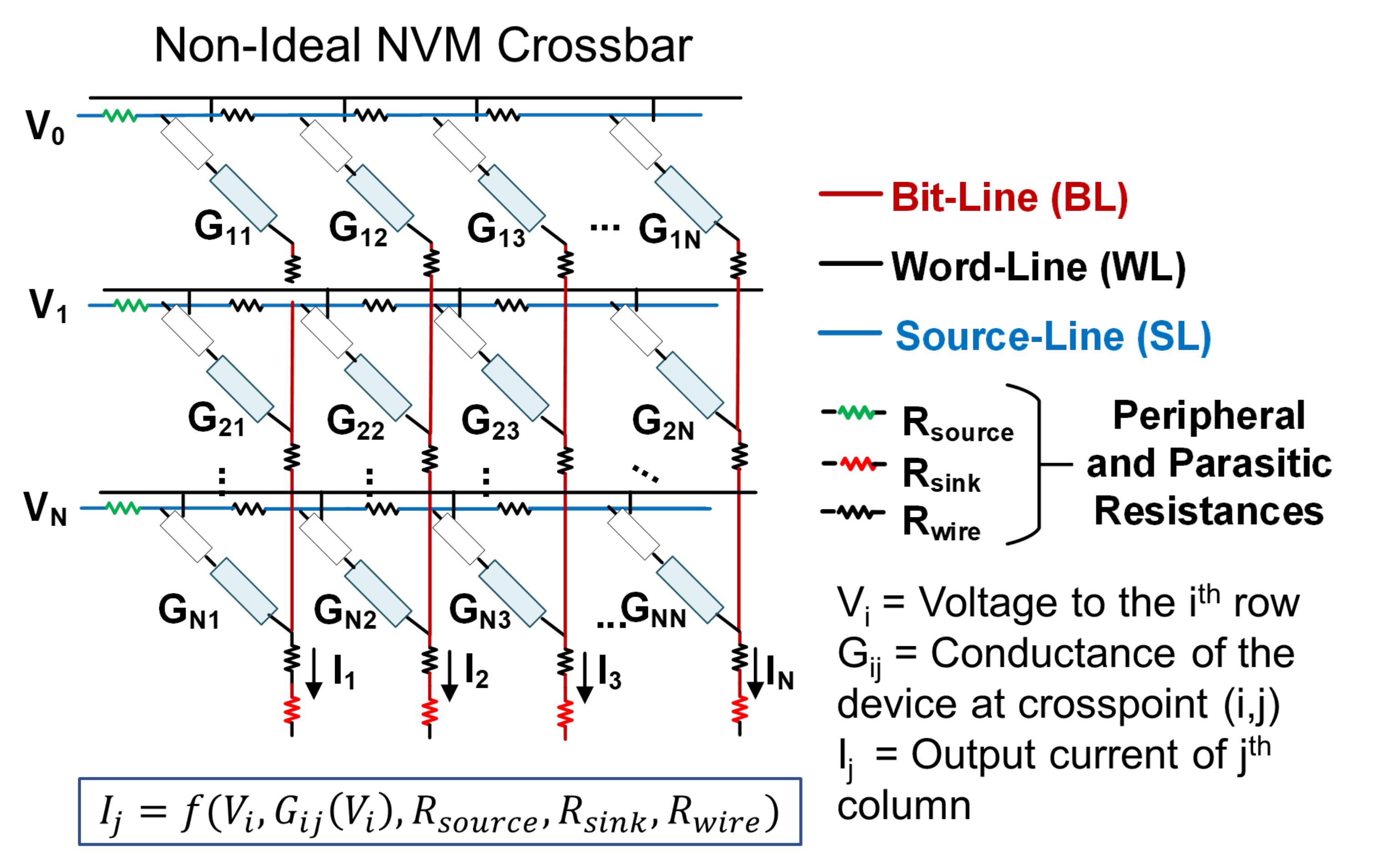}
	\caption{Illustration of NVM crossbar with various peripheral and parasitic resistances, which produces output current $I_j$, as a dot-product of voltage, $V_i$ and device conductance, $G_{ij}$}
	\label{fig:crossbar}

\end{figure}

The basic compute fabric in NVM technologies is a two-dimensional cross-point memory, known as a crossbar, shown in Fig. \ref{fig:crossbar}. The memory devices lie at the intersection of horizontally (source-line) and vertically (bit-line) running metal lines. The conductance of each memory device can be programmed to a discrete number of levels \cite{hu2016dac}. By simultaneously applying inputs, in the form of voltages, $V_i$, at the source-lines, the multiplications are performed between the voltages, $V_i$ and conductances, $G_{ij}$, by each NVM device using the principle of Ohm's law. Finally, the product, which is the resulting current, $I_{ij}$, from each NVM device, is summed up using Kirchoff's laws to produce a dot-product output, $I_j$ at each column:
\begin{equation}
    I_j =\sum_i I_{ij} =\sum_i V_iG_{ij}
\end{equation}




This analog nature of computing introduces errors in the MVM computations due to several non-idealities arising from the NVM devices and peripheral and parasitic resistances such as $R_{source}$, $R_{sink}$, $R_{wire}$, as shown in Fig. \ref{fig:crossbar}. Crossbar parameters, such as Crossbar Size, and ON Resistance have varying impact on the degree of functional errors introduced by the non-idealities \cite{chakraborty2020geniex} as they alter the effective resistance of a crossbar column. Larger crossbar size lowers the effective resistance, making the crossbar more prone to non-ideal effects, while higher ON resistance increases it, resulting in a crossbar less affected by non-idealities. 
 
Due to the non-idealities, the resulting output current, $I_{ni}$ is a function of voltage vector $V$, conductance matrix $G (V)$, which is now dependent on $V$, and several non-ideal factors:
\begin{equation}
    I_{ni} = f(V, G (V), R_{source}, R_{sink}, R_{wire})
    \label{eq:non-ideal}
    \ref{eq:non-ideal}
\end{equation}


We use GENIEx crossbar modeling technique \cite{chakraborty2020geniex}, which has a 2 layer perceptron network to model Equation \ref{eq:non-ideal}. The network predicts the output current, $I_{ni}$, from the crossbar for different input voltages, $V$ and conductances, $G$ of the NVM crossbar. The GENIEx model is trained using output current data obtained from HSPICE simulations of NVM crossbars considering all aforementioned non-idealities with sample $V$ and $G$ vectors and matrices, respectively.  



In order to evaluate DNNs on NVM crossbars in the presence of non-idealities, we use a simulation framework \cite{chakraborty2020geniex} following the standard technique of mapping convolutional and linear layers on a spatial NVM crossbar architecture such as PUMA \cite{ankit2019puma}. This mapping is composed of three parts: i) Iterative matrix-vector multiplications, ii) Tiling and iii) Bit-slicing. First, the convolution or linear layer operation in a neural network, is divided into iterative matrix vector computation. 
Second, the weight matrix is tiled into a number of crossbar sized segments. Third, since NVM devices can only accommodate limited number of bits, to represent larger bit precision inputs and weights, we perform bit-slicing. Here inputs and weights are divided into smaller bit portions, namely input streams and weight slices. 
Based on the crossbar size, device properties and bit-width of streams and slices, the crossbar model is obtained using the aforementioned GENIEx technique.  
The NVM crossbar non-idealities cause the activations at every layer to deviate from their expected value, and this deviation propagates through the network, resulting in  degradation of DNN accuracy at inference (without any adversary). Interestingly, the same deviation in activation imparts adversarial robustness when under attack, which is further analyzed in this paper.

\subsection{Adversarial Attacks}
 
In 2013, it was first demonstrated that a classifier can be forced to make an error by adding small perturbations to the data which are almost imperceptible to the human eye \cite{szegedy2013intriguing}. The term ``adversarial examples" was coined to define such data designed specifically to fool the classifier. Since then, several methods have been developed to generate such data, which are known as ``adversarial attacks". In principle, these attacks try to solve the following optimization problem \cite{papernot2017practical}:
\begin{equation}
\label{eq:untargetedAttack}
    x^{*} = x + argmin\{z: F(\theta, x + z)\neq F(\theta, x)\} = x + \delta_{x}
\end{equation}

where $x$ is the original data, $x^{*}$ is the perturbed adversarial data, $\theta$ is the model parameter, $F(\theta, x)$ is the classifier function, mapping inputs to labels, and the objective of the adversary is to misclassify, i.e. $F(\theta, x^{*}) \neq F(\theta, x)$. Most attacks use gradient-based optimization to solve for eq.\ref{eq:untargetedAttack}, and the attack's success relies on how accurately one can estimate $\nabla_{x}L(\theta, x, y)$, the derivative of the cost function $L(\theta, x, y)$ with respect to $x$ \cite{goodfellow2014explaining}.

\section{Adversarial Robustness of NVM Crossbar based Analog Computing}

In recent years, several adversarial defenses have been proposed that disrupt the gradient computation of the DNN by adding an extra computational element to the network, such as a randomization layer at the beginning \cite{xie2017mitigating}, or adaptive dropout after every layer \cite{dhillon2018stochastic}. When a DNN is implemented on an NVM crossbar architecture, the non-idealities have a similar effect of changing the layer-wise activations of the DNNs. There is no simple differentiable function to model these deviations, and one  cannot determine them without probing the analog hardware. Thus, such an implementation, could potentially increase the robustness of the neural network. In this section we describe the methodology to emulate DNNs on the PUMA architecture, and set up different threat scenarios based on the attacker's knowledge of both the software and the hardware. 

\subsection{Crossbar Models}
\label{sec:crossbar_models}

\begin{table}[h]
\centering
\vspace{-5pt}
\begin{tabular}{cccc}
\toprule
& \multicolumn{3}{c}{Crossbar parameters} \\ 
\cmidrule(r){2-4}
Crossbar Model & Size   & $R_{ON}$ ($\Omega$)  & $NF$ \\
\cmidrule(r){1-4}
64$\times$64\_300k  & 64$\times$64    & 300k    &     0.07  \\
32$\times$32\_100k  & 32$\times$32    & 100k    &     0.14  \\ 
64$\times$64\_100k  & 64$\times$64    & 100k    &     0.26  \\
\bottomrule
\end{tabular}
\vspace{-5pt}
\caption{Crossbar Model Description}
\label{tab:params}
\end{table}

To model the non-ideal crossbar, we use GENIEx \cite{chakraborty2020geniex}, as described in the previous section. For this work, we have replicated the modeling technique of GENIEx to generate 3 crossbar models (Table \ref{tab:params}). We used the RRAM device model \cite{guan2012spice} as the NVM device.

The degree of non-ideality is described in \cite{chakraborty2020geniex} as Non-ideality Factor ($NF$) $= Avg[ (Ideal\_Output- NonIdeal\_Output)/ Ideal\_Output]$. NF is directly (inversely) proportional to crossbar size (ON Resistance). In our experiments, we have considered different crossbar models to study the impact of different degrees of non-idealities, represented by different NF, on adversarial robustness, as shown in Table \ref{tab:params}. 
To implement this, we train different GENIEx crossbar models by creating datasets from data obtained by performing circuit simulations on the crossbar types listed in Table \ref{tab:params}. To integrate these NVM crossbar models with the PyTorch framework, we adopted the aforementioned PUMA functional simulator from \cite{chakraborty2020geniex} based on PUMA hardware architecture \cite{ankit2019puma}. 

\subsection{Datasets and Network Models}

For our evaluation we selected 3 image recognition tasks, and trained a ResNet \cite{he2016deep} for each task.
\begin{itemize}
    \item \textbf{CIFAR-10} \cite{krizhevsky2009learning}: A ResNet-20 was trained for 200 epochs and achieved test accuracy of $92.44\%$
    \item \textbf{CIFAR-100} \cite{krizhevsky2009learning}: A ResNet-32 was trained for 200 epochs and achieved test accuracy of $71.42\%$
    \item \textbf{ImageNet} \cite{deng2009imagenet}: A ResNet-18 was trained for 90 epochs and achieved top-1 test accuracy of $69.83\%$. We used a reduced test set of 1000 images for adversarial attacks. 
\end{itemize}

\subsection{Generating Adversarial attacks}

We define 4 different threat scenarios with varying extent of the attacker's knowledge of the target DNN and the underlying hardware (Table \ref{table:threat_models}). For each scenario, we defined an attack model (a DNN or an ensemble of DNNs) to  generate the adversarial images. We use Projected Gradient Descent (PGD)~\cite{madry2017towards} to generate iterative perturbations that are bound by the $l_{\infty}$ norm, as shown in Eq.\ref{eq:PGD}:
\begin{equation}
\label{eq:PGD}
    x^{t+1} = \Pi_{x+S}(x^{t} + \alpha sgn(\nabla_{x}L(\theta, x^{t}, y))
\end{equation}
$x^{t+1}$ is the adversarial example generated at $(t+1)^{th}$ iteration. The model's cost function is $L(\theta, x, y)$, which is a function of the model parameters $\theta$, input $x$, and labels $y$. The set of allowed perturbations is given by $S$. For the $l_{\infty}$ norm, the attack epsilon ($\epsilon$) defines the set of perturbations as $S = \Big(\delta | \big(x + \delta \geq max(x + \epsilon, 0)\big) \land \big(x + \delta \leq min(x + \epsilon, 1)\big)\Big) $, where $x \in [0, 1]$.

Additionally, for the two threat scenarios, non-adaptive, and adaptive Black Box attacks we also generated adversarial images using Square Attack \cite{ACFH2020square}, which is a query efficient adversarial Black Box attack. While PGD attack success is dependent on estimating the local gradients of the defending model, a query based attack doesn't rely on gradient information at all. Instead, it generates adversarial images by conducting a randomized search \cite{schumer1968adaptive, moonICML19}. Every time the attacker queries the model, the input image has random perturbations, sampled from a given distribution. If the perturbation succeeds in increasing the loss for that image, the image is updated, and this continues till either the image is misclassified, or the query limit is reached. 

\subsubsection{Non-Adaptive Attacks}

For non-adaptive attacks, we assume the attacker has no knowledge of the underlying analog hardware and the attacks are generated under the assumption of accurate digital computation. Under this category, we have devised 3 attacks.. 


\paragraph{Ensemble Black Box Attack:} The attacker queries the model on an accurate digital hardware and reads the output of the final layer before softmax (logits) to generate a synthetic dataset of training data and its corresponding logits. This synthetic dataset is used to train 3 different surrogate ResNet models, ResNet-10,20,32. These 3 models are then used to generate adversarial images using the stack parallel ensemble strategy \cite{hang2020ensemble}. 

\paragraph{Square Attack (Black Box):} The attacker queries the model on accurate digital hardware and has access to the last layer (logits) as in the case of the Ensemble Black Box Attacks. We use $l_{\infty}$ Square Attack \cite{ACFH2020square}, and set the maximum query limit to 1000. 

\paragraph{White Box Attack:} This is the highest threat level where the attacker has full knowledge of the model weight, thus the attack model is the same as the target model. However, while generating gradients, the attacker has no knowledge of the underlying analog hardware implementation. The gradients for the attack are computed assuming accurate digital hardware implementation.

\subsubsection{Hardware in Loop Adaptive Attacks}

The attacker is aware that the model is implemented on an NVM crossbar hardware. However, the crossbar model available to the attacker may or may not match with the target's implementation due to different hardware technologies
For crafting Ensemble Black Box Attacks, the attacker queries the DNN model implemented on the NVM crossbar based hardware to create the synthetic dataset. Similarly, for Square Attack, the repeated queries are made for the DNN implemented on the Analog Hardware. As emulation of the crossbar based architecture take much longer, we limit the total number of queries to 30. In the case of White Box attacks, the attacker generates adversarial images using ``Hardware-in-Loop" gradient descent. Note that the NVM crossbar based hardware is designed for inference tasks and does not support backpropagation of gradients. Thus, for  ``Hardware-in-Loop", the forward pass is performed on NVM crossbar hardware, and all activations are recorded. However, the derivatives are calculated assuming ideal computations in place of non-ideal MVM operations of the crossbar. As described in Section \ref{sec:crossbar_models}, the NVM crossbar non-idealities vary with crossbar properties. We use 3 different crossbar models as defined in Table \ref{tab:params} and we explore scenarios where there is a mismatch in the crossbar model used by the attacker and the target implementation.

\subsubsection{Comparison with Related Work}
We have selected 3 defenses that can be applied to a pretrained network as listed below. For a fair comparison, we apply non-adaptive attacks for these defenses as well, i.e. the defenses are not visible to the attacker when they query the model for the two Black Box attacks, and when they generate gradients for White Box attack.

\begin{itemize}
     \item \textit{Input Bit Width (BW) Reduction} \cite{guo2017countering}: The input is quantized to 4-bits.
    \item \textit{Stochastic Activation Pruning (SAP)} \cite{dhillon2018stochastic} (for CIFAR-10/100 only): At inference, after every convolution layer, there is an adaptive dropout, that randomly sets the layer outputs to 0 with a probability proportional to their absolute value.
    \item \textit{Random Padding} \cite{xie2017mitigating} (for ImageNet only): Two randomization layers are introduced before the pretrained model. The first layer scales the input image to a random size NxN where N $\in$ [299, 331] using nearest-neighbor extrapolation. The second layer randomly pads the image to generate the final image of size 331x331.
\end{itemize}

\section{Results}
\begin{table*}[h]
\vspace{-5pt}
\resizebox{0.8\textwidth}{!}{
\begin{tabular}{ccccccc}
\toprule
 &               & \multicolumn{2}{c}{Accurate Digital Computation} & \multicolumn{3}{c}{Non-Ideal Analog Computation} \\
\cmidrule(r){3-4} \cmidrule{5-7}
 Attack Type & Model Weights & Logits & Activations & Crossbar Model   & Logits  & Activations  \\
 \cmidrule(r){1-7}
Non-Adaptive Attacks     &     &     &     &  &     &     \\
\cmidrule(r){1-1}

Black Box Attacks    & No  & Yes & No  & No                  & No  & No  \\
White Box Attacks    & Yes & Yes & Yes & No                  & No  & No  \\
\cmidrule{1-7}
Adaptive Attacks     &     &     &     &  &     &     \\
\cmidrule(r){1-1}
Black Box Attacks    & No  & N/A & N/A & Yes (may not match) & Yes & No  \\
White Box Attacks    & Yes & N/A & N/A & Yes (may not match) & Yes & Yes \\
\bottomrule
\end{tabular}}
\vspace{-5pt}
\caption{Attacker's Knowledge for the Threat Scenarios}
\label{table:threat_models}
\end{table*}

The first effect of implementing DNNs on a NVM crossbar hardware is the reduction in clean accuracy due to the errors associated with non-ideal computations. Greater the Non-Ideality Factor (NF), more severe is the accuracy degradation as noted in Table \ref{table:non_adaptive}. The clean accuracy of CIFAR-10 drops from 92.44\% (accurate digital hardware) to 88.34\% on 64x64\_100k, the most non-ideal crossbar model among the three. Similarly, CIFAR-100 accuracy drops from 71.42\% to 55.48\% and ImageNet accuracy falls from 69.56\% to 62.50\%. If non-idealities of NVM hardware had no impact on adversarial robustness, similar degradation would have been observed in DNN accuracy under attack. However, our findings, as outlined below, indicate a different trend. 

\subsection{Non-Adaptive Attacks}

\begin{figure}[h]
\centering
\subcaptionbox{\label{fig:bbox_c10}}{\includegraphics[width=0.43\linewidth]{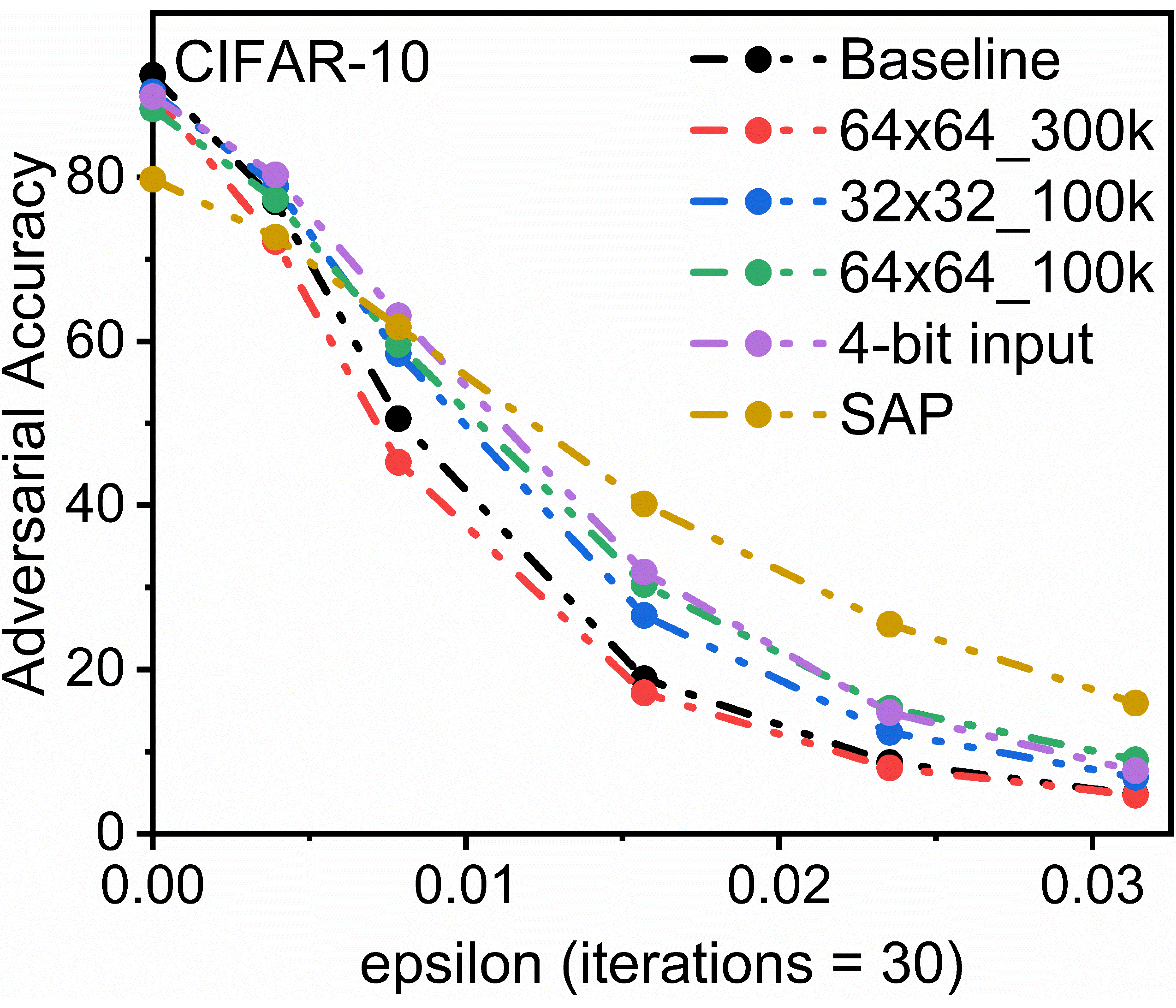}}
\hfil
\subcaptionbox{\label{fig:bbox_c100}}{\includegraphics[width=0.43\linewidth]{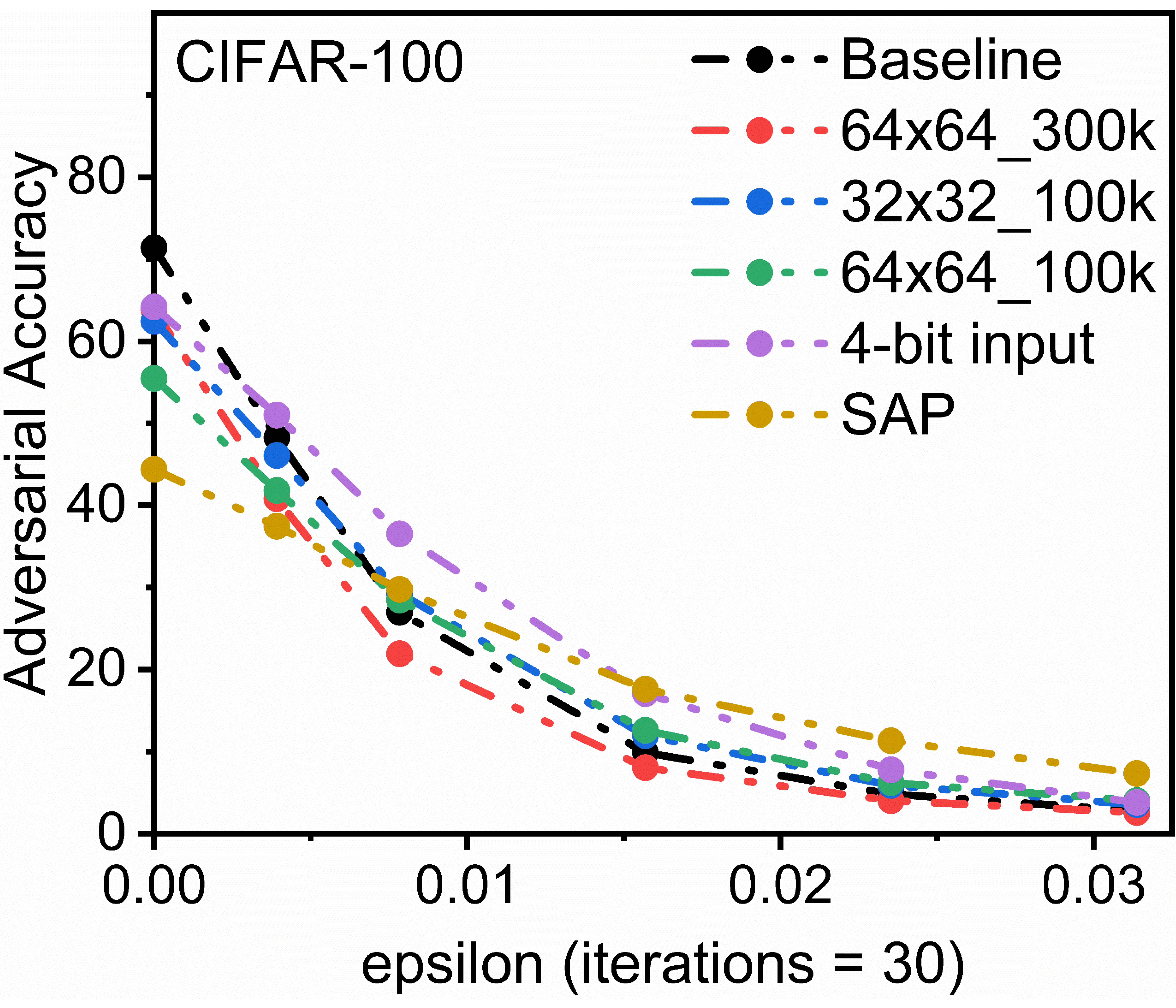}} 
\caption{Non-Adaptive Ensemble (Black Box) PGD (iter=30) on CIFAR-10, CIFAR-100 on 3 NVM crossbar models and the 2 defenses, Input BW Reduction (4-bit input) \cite{guo2017countering} and SAP \cite{dhillon2018stochastic}}
\label{fig:bbox_attacks}
\end{figure}

\paragraph{Ensemble Black Box Attacks} From Fig.\ref{fig:bbox_attacks}, we observe the decline in adversarial accuracy with increasing attack epsilon ($\epsilon$) for CIFAR-10/100. The 64x64\_300k model didn't exhibit any increase in robustness, instead it trailed behind the baseline accuracy. The NVM crossbar models, 32x32\_100k and 64x64\_100k, recorded an absolute increase in robustness of 5.3\% and 7.8\% averaged over $\epsilon$ = (2,4,6,8)/255, respectively for CIFAR-10. For CIFAR-100, it was 1.4\% and 1.84\% respectively. The peak improvement in robustness was observed for $\epsilon$ = 4/255 and has been summarized in Table \ref{table:non_adaptive}.

\begin{figure}[h]
\centering
\subcaptionbox{\label{fig:sqbbox_c10}}{\includegraphics[width=0.43\linewidth]{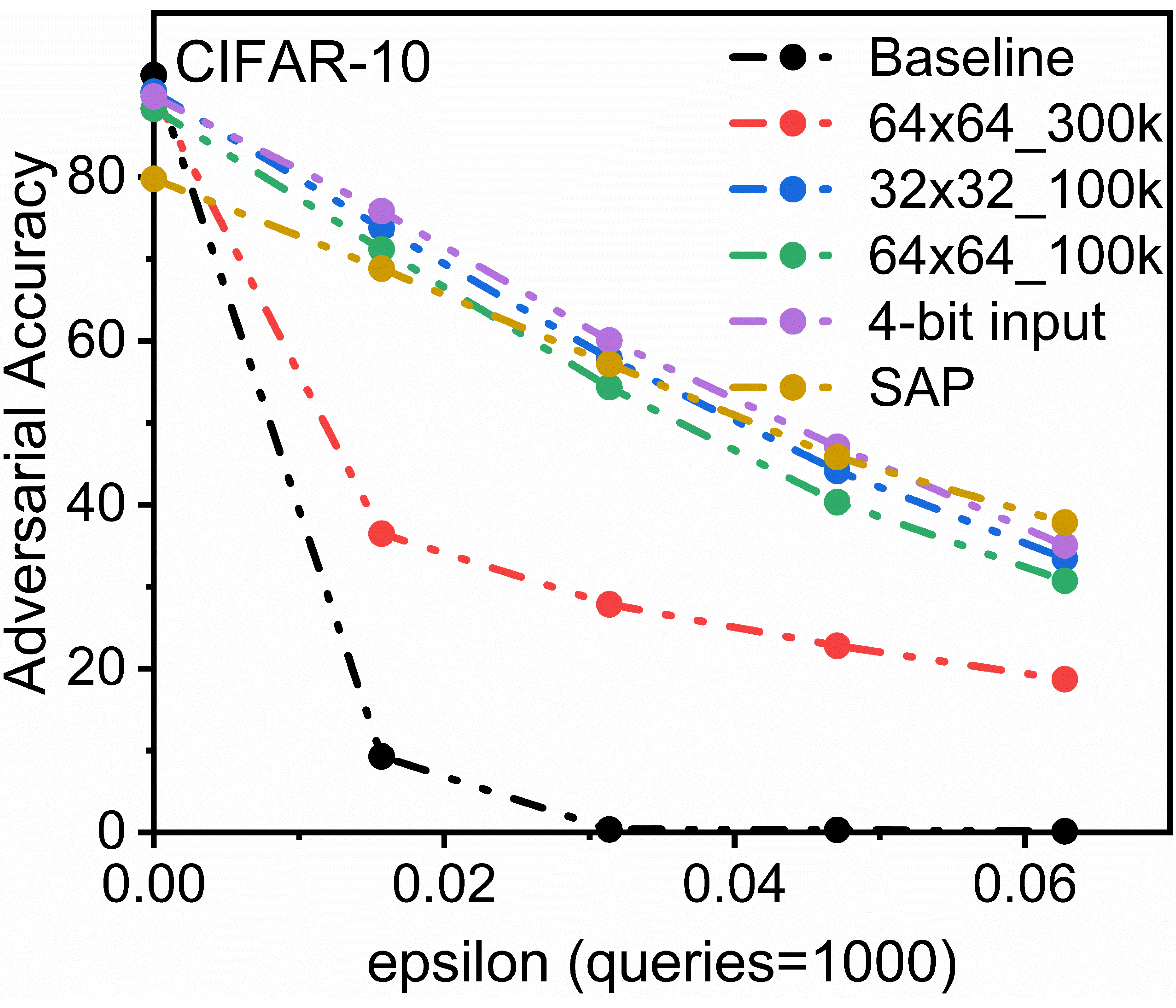}} 
\hfil
\subcaptionbox{\label{fig:sqbbox_c100}}{
\includegraphics[width=0.43\linewidth]{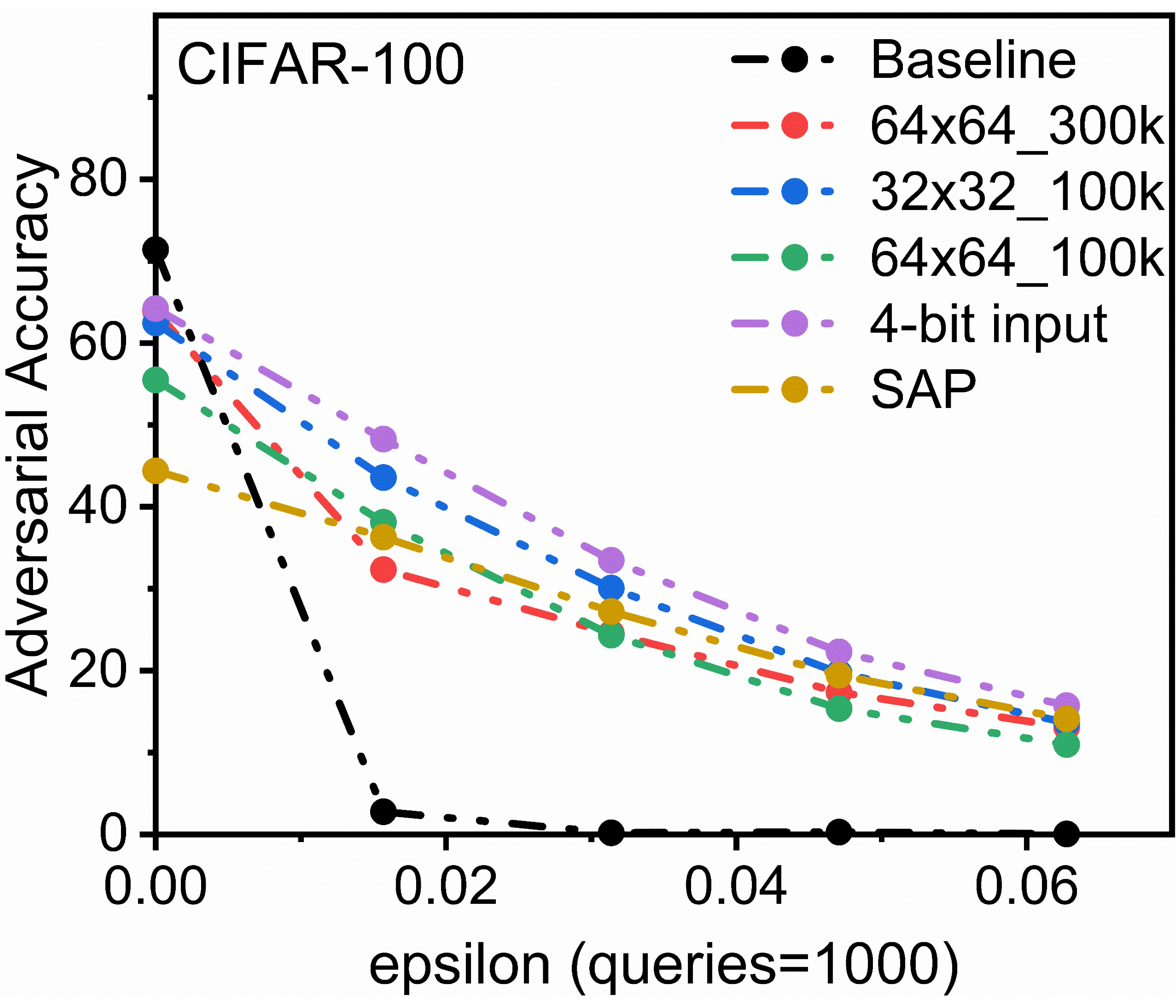}}
\hfil
\subcaptionbox{\label{fig:sqbbox_imnet}}{ \includegraphics[width=0.43\linewidth]{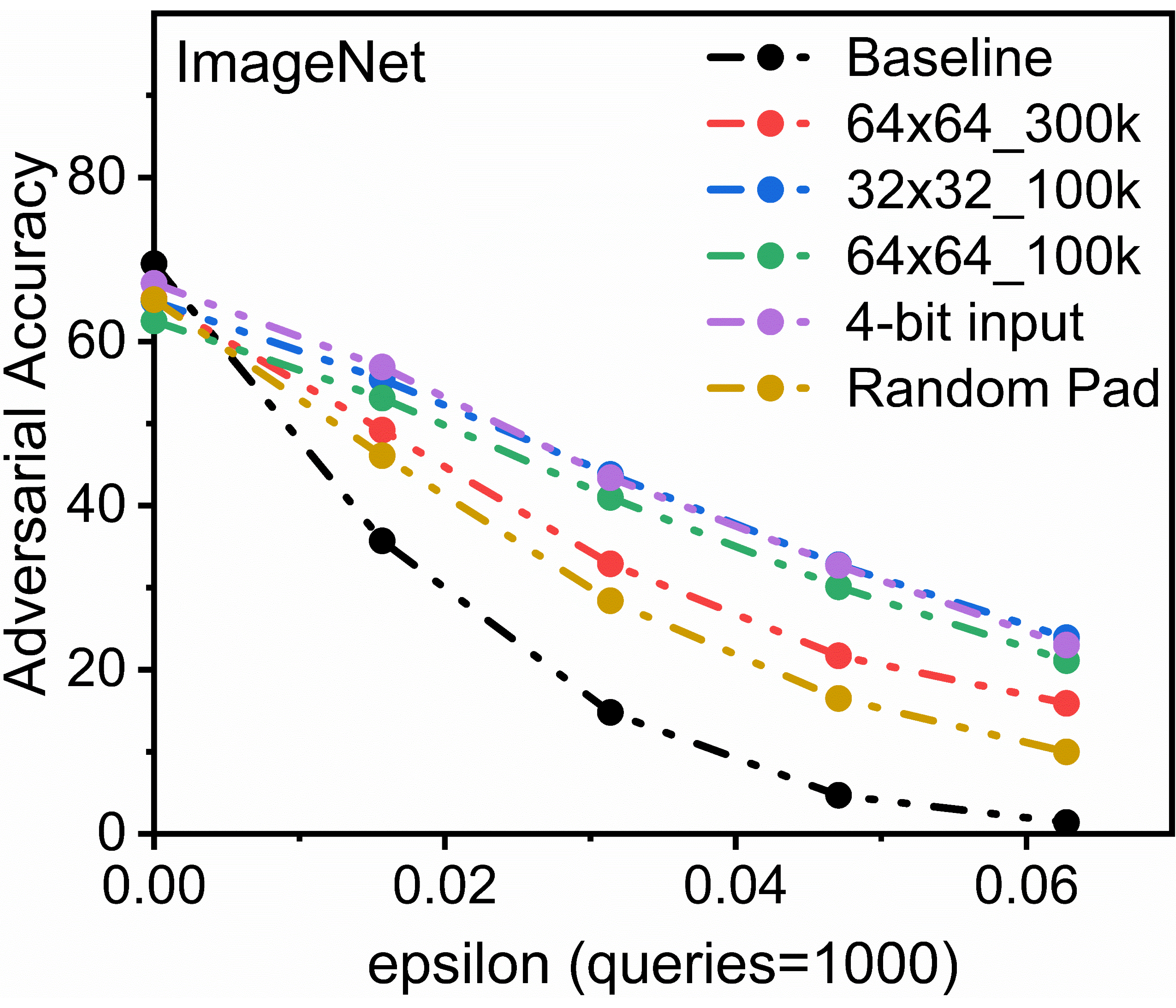}} 
\caption{Non-Adaptive Square Attacks (Black Box) on CIFAR-10/100 and ImageNet on 3 NVM models and 3 defenses, Input BW Reduction (4-bit input) \cite{guo2017countering}, SAP \cite{dhillon2018stochastic}, Random Pad \cite{xie2017mitigating}}
\label{fig:sqbbox_attacks}
\end{figure}

\paragraph{Square Attack (Black Box)} The analog hardware shows the highest resilience against such an attack. As the attack is gradient-free in nature, we conclude that analog hardware offers robustness by modifying the inference itself. The perturbations that cause complete DNN failure, i.e. 0\% accuracy, have much lower impact on the DNN implemented on NVM crossbar. The other 3 defense methods \cite{guo2017countering, dhillon2018stochastic, xie2017mitigating}, also perform well over a wide range of $\epsilon$=(4,8,12,16)/255. The average robustness gain observed for CIFAR-10 was 23.93\% , 49.80\% and 46.63\% with crossbar models as 64x64\_300k, 32x32\_100k, 64x64\_100k  respectively. We see robustness gain increase from 64x64\_300k to 32x32\_100k, and then drop slightly for 64x64\_100k. The increase if due to higher deviations in 32x32\_100k compared to 64x64\_300k. The slight decrease however, can be attributed to the counter effect of inaccurate computations as non-idealities increase further. We observe similar trends in CIFAR-100 and Imagenet as well, as shown in Fig. \ref{fig:sqbbox_attacks} and Table \ref{table:non_adaptive}.

\begin{figure}[h]
\centering
\subcaptionbox{\label{fig:wbox_c10}}{ \includegraphics[width=0.43\linewidth]{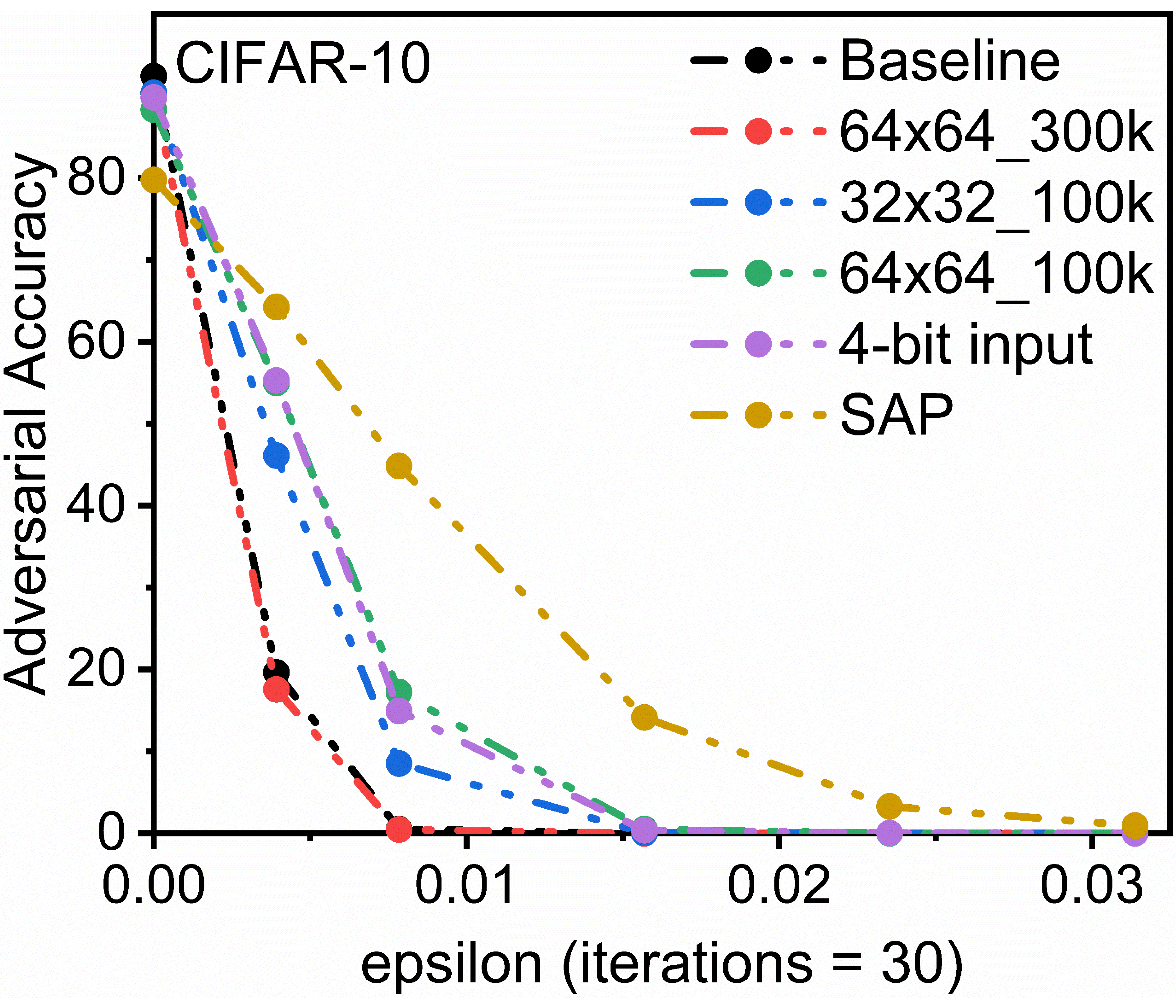}}
\hfil
\subcaptionbox{\label{fig:wbox_c100}}{\includegraphics[width=0.43\linewidth]{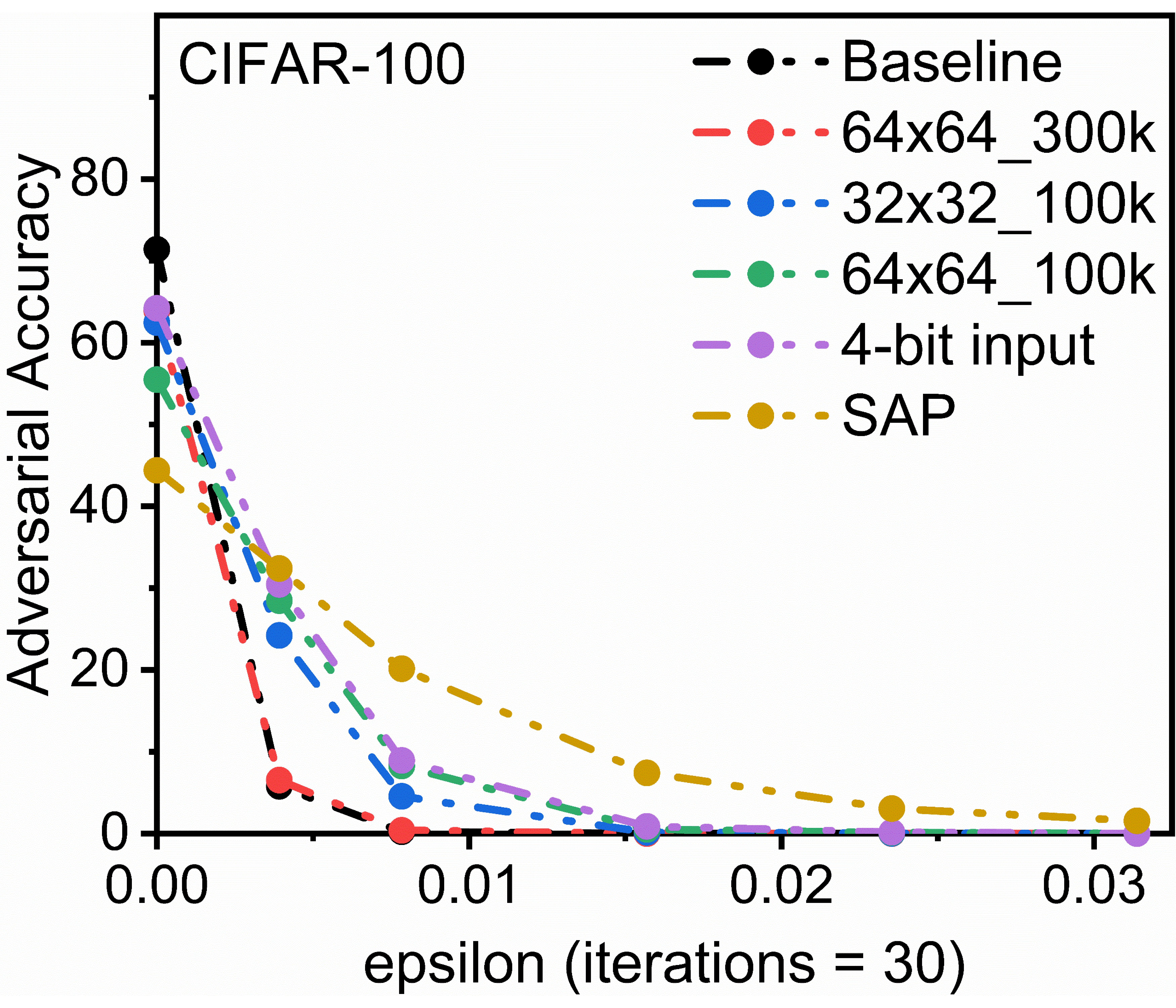}}
\caption{Non-Adaptive White Box Attacks (PGD, iter=30) on CIFAR-10, CIFAR-100 on 3 NVM models and 2 defenses, Input BW Reduction (4-bit input) \cite{guo2017countering} and SAP \cite{dhillon2018stochastic}}
\label{fig:wbox_attacks}
\end{figure}

\paragraph{White Box Attacks} Under this threat model we observe substantial improvement in robustness as depicted in Fig. \ref{fig:wbox_attacks} and Table \ref{table:non_adaptive}. The NVM model 64x64\_300k still continues to closely follow baseline accuracy. For all 3 datasets, the baseline accuracy drops sharply to 0 beyond $\epsilon$ = 2/255.
At this level, the NVM models are no longer able to recover any performance. We observe that 64x64\_100k, the most non-ideal of the 3 models, offers the highest improvement for all 3 datasets, with absolute increase of 35.34\% for CIFAR-10, 22.69\% for CIFAR-100, and 9.90\%  for ImageNet at $\epsilon$ = 1/255.

\begin{table*}[h]
\centering
\caption{Summary of Non-Adaptive Attacks on NVM Crossbar Models}
\vspace{-5pt}
\label{table:non_adaptive}
\resizebox{\textwidth}{!}{
\begin{tabular}{lrrrrrr}
\toprule
 &  & \multicolumn{3}{c}{NVM Crossbar Models (Target)} & \multicolumn{2}{c}{Related Work} \\
 \cmidrule(r){3-5} \cmidrule(r){6-7}
Attack Type & \multicolumn{1}{c}{Baseline} & \multicolumn{1}{c}{64$\times$64\_300k} & \multicolumn{1}{c}{32$\times$32\_100k} & \multicolumn{1}{c}{64$\times$64\_100k} & \multicolumn{1}{c}{4-bit input \cite{guo2017countering}} & \multicolumn{1}{c}{SAP \cite{dhillon2018stochastic}} \\
\cmidrule(r){1-7}
& \multicolumn{6}{c}{CIFAR-10 (ResNet-20) (test samples = 10000)} \\
\cmidrule(r){2-7}
Clean   
& 92.44 & 90.35 (-2.09) & 90.42 (+2.02)  & 88.34 (-4.10)  & 89.84 (-2.60)  & 79.76 (-12.68) \\
Ensemble (Black Box) PGD $\epsilon$ = 4/255, iter = 30  
& 18.91 & 17.15 (-1.76) & 26.6 (+7.69)  & 30.35 (+11.44) & 31.89 (+12.98) & 40.19 (+21.28) \\
Square Attack (Black Box) $\epsilon$ = 4/255, queries = 1000
& 9.29 & 36.47 (+27.18) & 73.79 (+64.50) & 71.18 (+61.89) & 75.85 (+66.56) & 68.84 (+59.55) \\
White Box PGD $\epsilon$=1/255, iter = 30
& 19.64 & 17.56 (-2.08) & 46.12 (+26.48) & 54.98 (+35.34) & 55.29 (+35.65) & 64.26 (+44.62) \\
White Box PGD $\epsilon$=2/255, iter = 30
& 0.51  & 0.45 (-0.06)  & 8.51 (+8.00)   & 17.22 (+16.71) & 14.94 (+14.34) & 44.85 (+44.34) \\
\cmidrule(r){2-7}
& \multicolumn{6}{c}{CIFAR-100 (ResNet-32) (test samples = 10000)}  \\
\cmidrule(r){2-7}
Clean 
& 71.42 & 63.89 (-7.53) & 62.44 (-8.98)  & 55.48 (-15.94) & 64.20 (-7.22)  & 44.41 (-27.01) \\
Ensemble (Black Box) PGD $\epsilon$=4/255, iter = 30
& 9.88  & 8.03 (-1.85)  & 11.95 (+2.07)  & 12.59 (+2.71)  & 17.07 (+7.19)  & 17.60 (+7.72)  \\
Square Attack (Black Box) $\epsilon$ = 4/255, queries = 1000
& 2.76 & 32.33 (+29.57) & 43.59(+40.83) & 38.12 (+35.36) & 48.28 (+45.52) & 35.25 (+32.49) \\
White Box PGD $\epsilon$=1/255, iter 30
& 5.78  & 6.53 (+0.75)  & 24.22 (+18.44) & 28.47 (+22.69) & 30.45 (+24.67) & 32.4 (+26.62)  \\
White Box PGD $\epsilon$=2/255, iter 30
& 0.24  & 0.39 (+0.15)  & 4.55 (+4.31)   & 8.27 (+8.03)   & 8.94 (+8.70)   & 20.14 (+19.9) \\
\cmidrule(r){2-7}
& \multicolumn{6}{c}{ImageNet (ResNet-18) (test samples = 1000)}  \\
\cmidrule(r){2-7}
 & \multicolumn{1}{c}{Baseline} & \multicolumn{1}{c}{64$\times$64\_300k} & \multicolumn{1}{c}{32$\times$32\_100k} & \multicolumn{1}{c}{64$\times$64\_100k} & \multicolumn{1}{c}{4-bit input \cite{guo2017countering}} & \multicolumn{1}{c}{Random Pad \cite{xie2017mitigating}} \\
\cmidrule(r){2-7} 
Clean  & 69.56   & 65.2 (-4.36) & 64.9 (-4.66) & 62.5 (-7.06) & 67.1 (-2.46)  & 65.1 (-4.46) \\
Square Attack (Black Box) $\epsilon$ = 4/255, queries = 500 
& 35.70 & 49.20 (+13.50) & 55.40 (+19.70) & 53.10 (+17.40) & 56.90 (21.20) & 46.10 (+10.40) \\
White Box PGD $\epsilon$=1/255 , iter = 30
& 0.40 & 0.60 (+0.20)  & 4.50 (+4.10) & 10.30 (+9.90) & 9.6 (+9.20) & 44.3 (+43.90) \\
White Box PGD $\epsilon$=2/255, iter = 30 
& 0.10  & 0.10 (+0.00)  & 0.20 (+0.10)  & 0.50 (+0.40) & 0.10 (+0.00)  & 33.50 (+33.40) \\
\bottomrule
\end{tabular}
}
\vspace{-5pt}
\end{table*}

\begin{figure}[h]
\centering
\subcaptionbox{
\label{fig:gain_vs_NF_c10}}{
\includegraphics[width=0.43\linewidth]{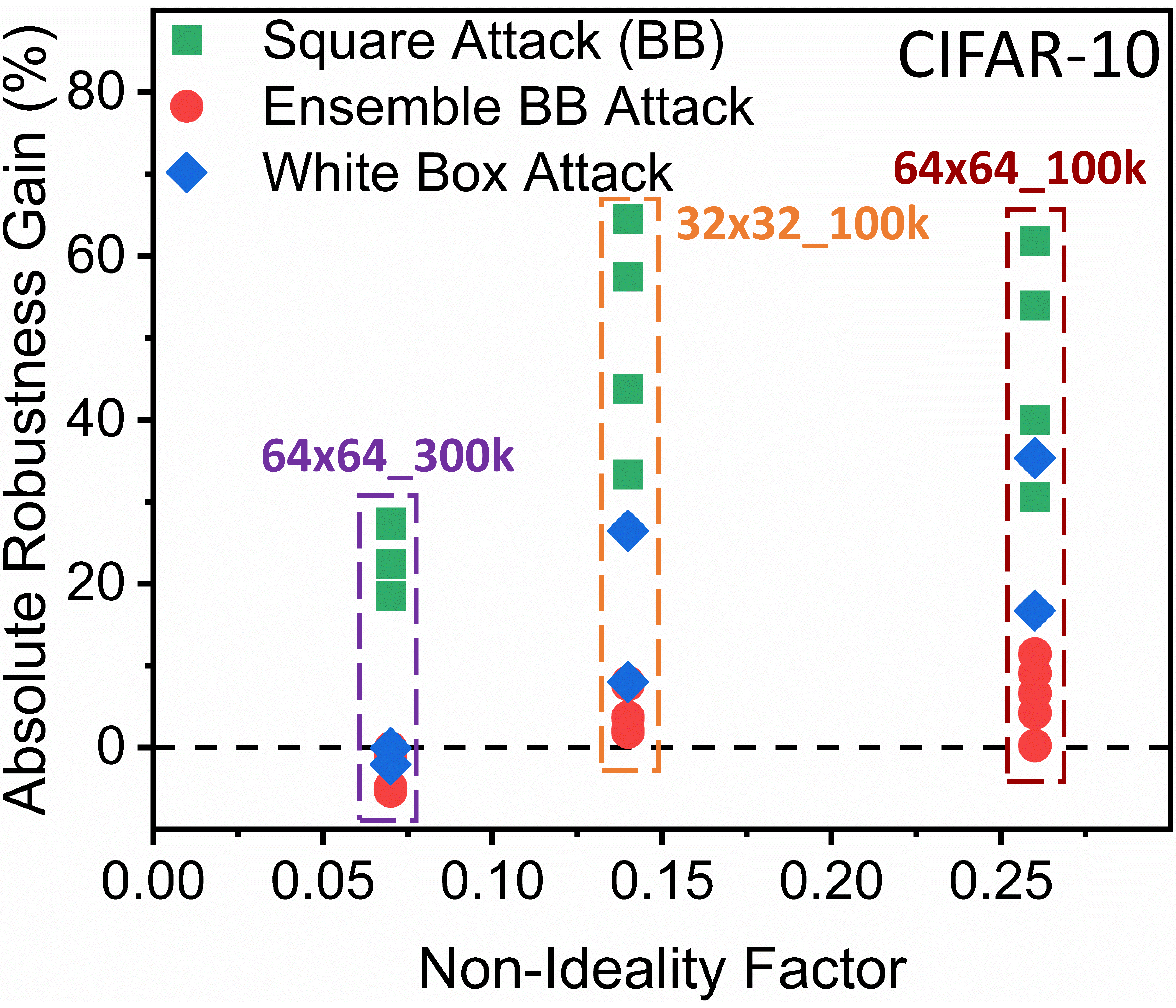}} \hfil
\subcaptionbox{\label{fig:gain_vs_NF_c100}}{
\includegraphics[width=0.43\linewidth]{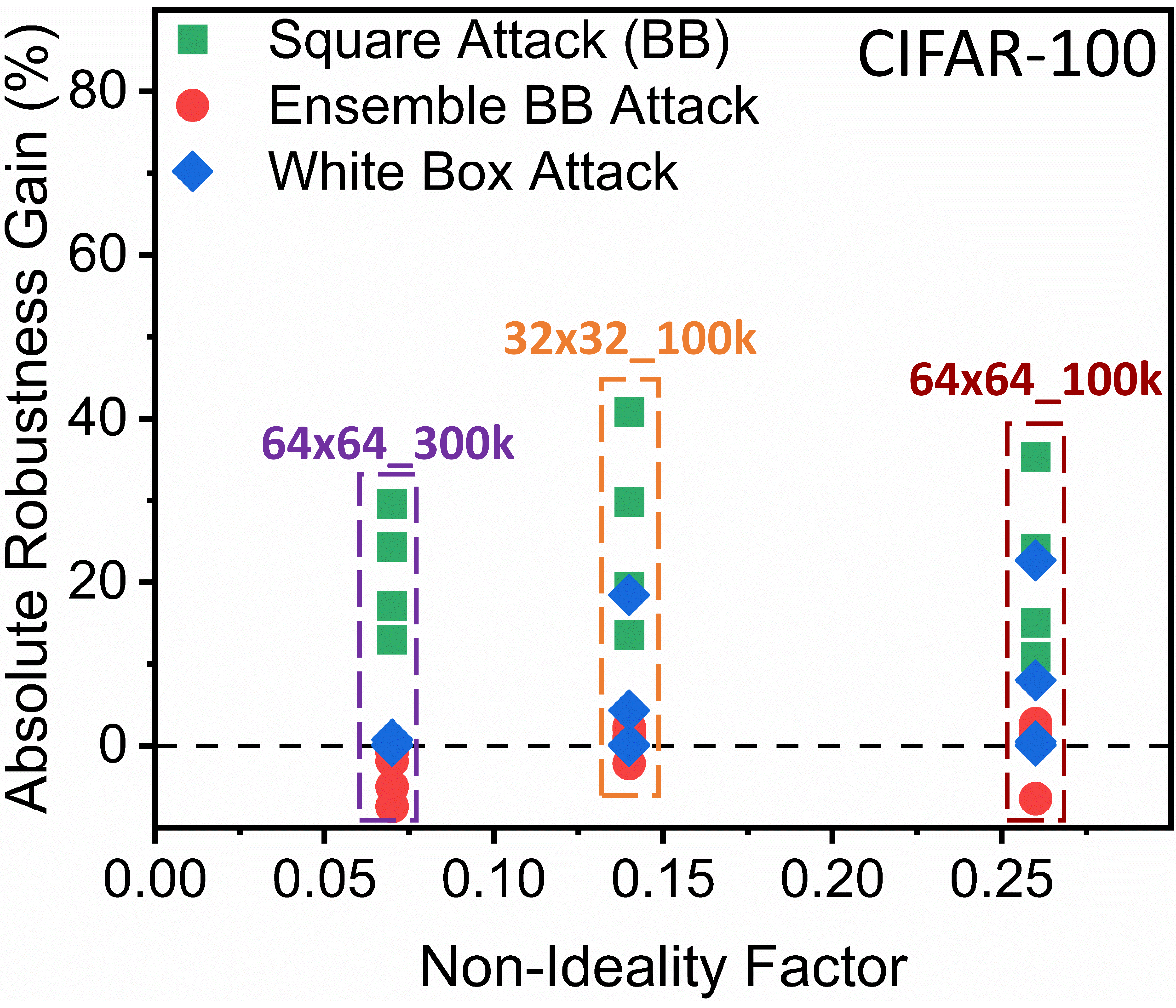}} 
\caption{Absolute Gain in Adversarial Accuracy for various adversarial attacks vs the Non-Ideality Factor of NVM Crossbars for CIFAR-10/100}
\label{fig:gain_vs_NF}
\vspace{-5pt}
\end{figure}

We have summarized below the trends observed across all non-adaptive attacks. 
\begin{itemize}
    \item For gradient based attacks (PGD) \cite{madry2017towards},
    more the attacker relies on estimating the true gradients of the target model for attack generation, greater is absolute robustness gain. We observed an increase in the absolute improvement from baseline accuracy as we move from Ensemble Black Box to White Box attacks.
    \item The resulting accuracy is a combination of two opposing forces. The errors caused by the non-idealities try to lower the accuracy, while the intrinsic robustness lowers the effectiveness of the attack and pushes the accuracy higher than the baseline. For example, for 64x64\_300k (NF = 0.07),  the MVM operations are closest to ideal computation for the non-adaptive attacks to transfer successfully. Whereas, the more non-ideal crossbar models, 32x32\_100k and 64x64\_100k, have greater clean accuracy degradation due to functional errors, but have higher adversarial accuracy, as the non-idealities hinder the transfer of the attacks. This push-pull effect can also be seen in Fig. \ref{fig:gain_vs_NF}, where we plot the robustness gain vs crossbar NF for all the non-adaptive attacks. We see a significant difference as NF increases from 64x64\_300k to 32x32\_100k. At 64x64\_100k, we see the gain taper slightly below 32x32\_100k, as inaccurate computations start to have a greater impact over intrinsic robustness. 
\end{itemize}

\subsection{Hardware-in-Loop Adaptive Attacks}

\begin{figure}[h]
\centering
\subcaptionbox
{\label{fig:bbox_c10_hwinloop}}{
\includegraphics[width=0.43\linewidth]{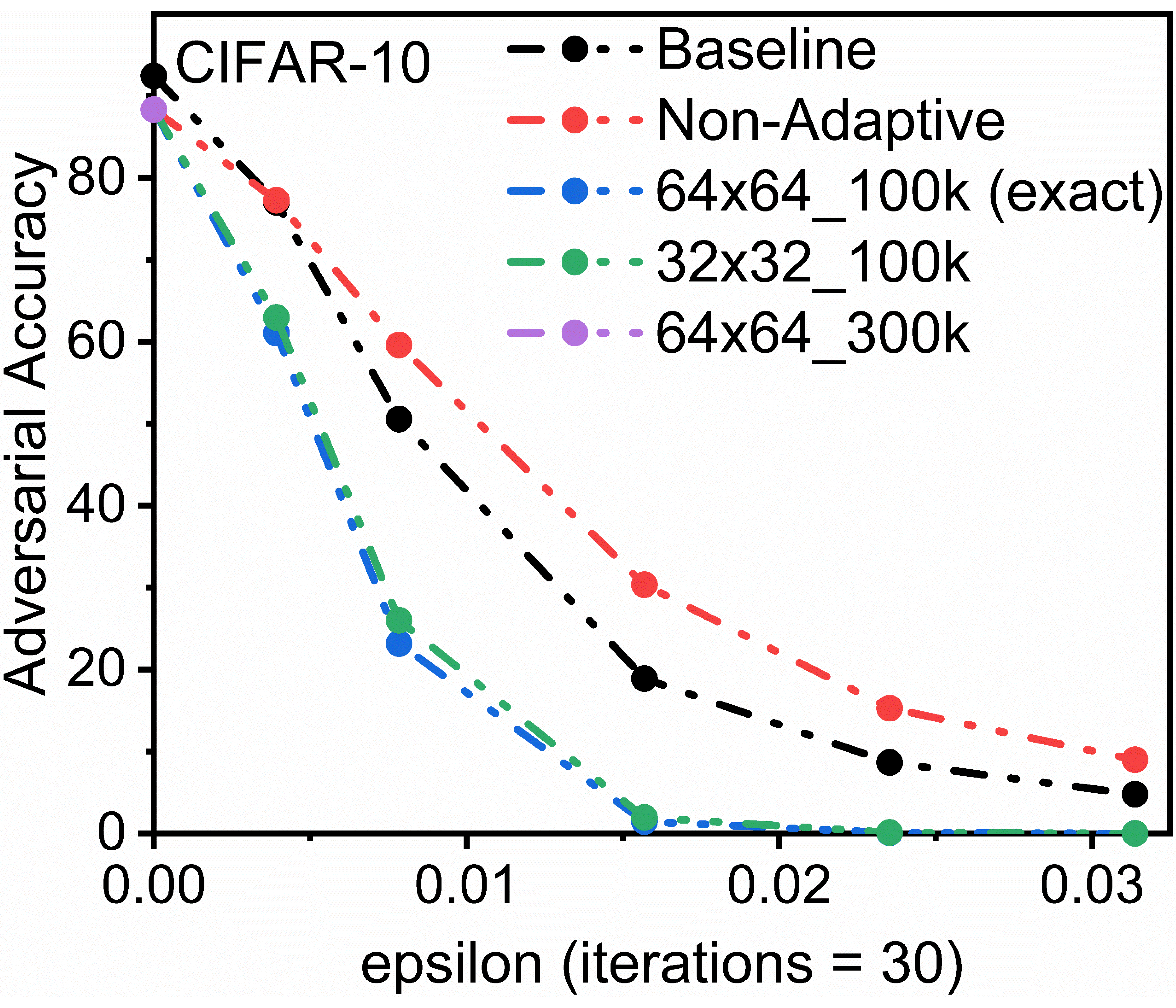}} \hfil
\subcaptionbox{\label{fig:bbox_c100_hwinloop}}{
\includegraphics[width=0.43\linewidth]{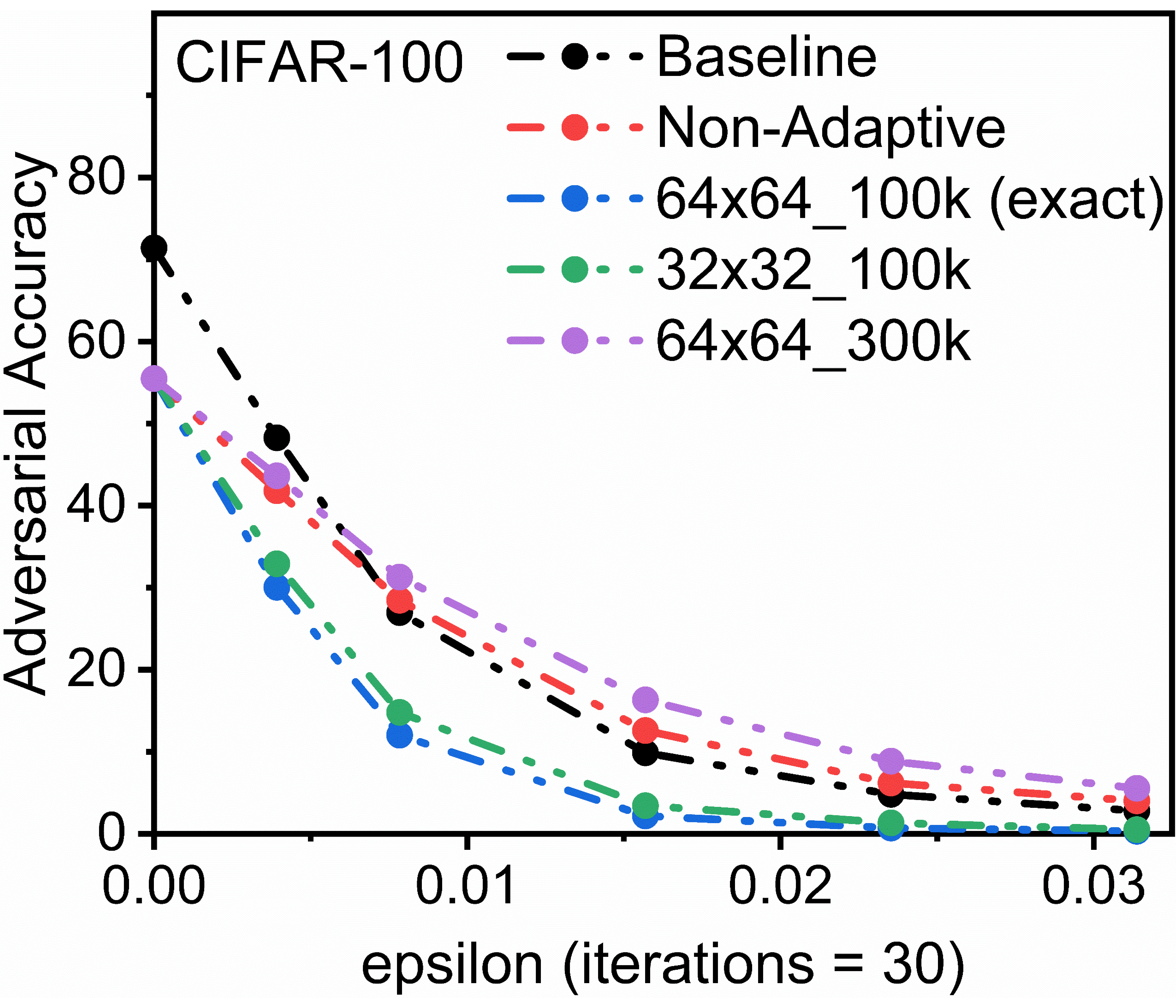}} 
\caption{Hardware-in-Loop Adaptive Black Box Attacks (PGD, iter=30) on CIFAR-10/100. Target NVM model is 64x64\_100k, and the attacks are generated using 3 different NVM models. }
\label{fig:bbox_attacks_hwinloop}
\vspace{-5pt}
\end{figure}

\begin{table*}[h!]
\centering
\caption{Hardware-in-Loop Adaptive Attacks: The values in bold indicate attacker's NVM model is a match to defender's NVM model}
\vspace{-5pt}
\label{table:adaptive}
\resizebox{0.7\textwidth}{!}{
\begin{tabular}{lccrrrr}
\toprule
& & & & \multicolumn{3}{c}{NVM Crossbar Model (Target)} \\
\cmidrule(r){5-7}
Dataset & Test Samples & Attack $\epsilon$ & \multicolumn{1}{c}{Baseline} & \multicolumn{1}{c}{64$\times$64\_300k} & \multicolumn{1}{c}{32$\times$32\_100k} & \multicolumn{1}{c}{64$\times$64\_100k} \\
\cmidrule{1-7}
\multicolumn{3}{c}{Ensemble BB Attack (iter=30)} & & \multicolumn{3}{c}{Attacker's NVM Crossbar model: 64$\times$64\_100k}   \\
\cmidrule{1-3}
\cmidrule{5-7}
CIFAR-10 & 10000 & 4/255 & 18.91 & 1.95 (-16.96) & 1.45 (-17.46) & \textbf{1.27} (-17.64) \\
CIFAR-100 & 10000 & 4/255 & 9.88 & 8.54 (-1.34) & 2.74 (-7.74) & \textbf{2.17} (-7.71) \\
\cmidrule{1-7}
\multicolumn{3}{c}{Square Attack (BB) (queries=30)} & & \multicolumn{3}{c}{Attacker's NVM Crossbar model: 32$\times$32\_100k}   \\
\cmidrule{1-3}
\cmidrule{5-7}
CIFAR-10 & 1000 & 8/255 & 67.50 & 71.80 (+4.30) &  \textbf{66.60} (-0.90) & 64.10 (-3.40) \\
CIFAR-100 & 1000 &  8/255 & 40.10 & 49.20 (+9.1) & \textbf{32.50} (-7.60) & 26.70 (-13.40)\\
Imagenet & 1000 & 8/255 & 48.50 & 53.30 (+4.80) & \textbf{46.00} (-2.50) & 44.30 (-4.20) \\
\cmidrule{1-3}
\cmidrule{5-7}
\multicolumn{3}{c}{White Box PGD (iter=30)} & & \multicolumn{3}{c}{Attacker's NVM Crossbar model: 64$\times$64\_100k}   \\
\cmidrule{1-3}
\cmidrule{5-7}
CIFAR-10 & 10000 & 1/255 & 19.64  & 43.45 (+23.81) & 31.78 (+12.14)  & \textbf{28.84} (+9.2) \\
CIFAR-10 & 10000 & 2/255 & 0.51   & 6.98 (+6.47)   & 2.13 (+1.62)     & \textbf{1.87} (+1.36) \\
CIFAR-100 & 10000 & 1/255 & 5.78  & 28.21 (+22.43) & 10.86 (+5.08) & \textbf{9.73} (+3.95) \\
ImageNet & 1000 & 1/255 & 0.40  & -- & -- & \textbf{0.80} (+0.40)  \\
\bottomrule
\end{tabular}
}
\vspace{-5pt}
\end{table*}

\paragraph{Ensemble Black Box Attacks} When the attacker builds their synthetic dataset by querying the NVM crossbar hardware implementation of the DNN, the resulting Ensemble Black Box attacks are much more effective. The adversarial accuracy of the hardware falls significantly below the baseline, as shown in Fig. \ref{fig:bbox_attacks_hwinloop} and Table \ref{table:adaptive}. Even when the attack is built using a crossbar model different from the target, accuracy degradation is significant. We observe that attacks generated using 32x32\_100k (NF = 0.14) are stronger than those generated using 64x64\_300k (NF = 0.07) when applied to 64x64\_100k (NF= 0.26). This implies that the lesser the difference in NF, the more effective are the attacks. 

\paragraph{Square Attack (Black Box)} By repeatedly querying the actual NVM crossbar based hardware, the attacker could generate much stronger attacks, as shown in Fig. \ref{fig:sqbbox_attacks} and Table \ref{table:adaptive}. In fact, the generated attacks are as strong as the baseline, however, when there is a significant mismatch in hardware properties, the attack doesn't transfer well.

\paragraph{White Box Attacks} 
Even when the attacker has full knowledge of the hardware, the non-idealities help improve robustness for White Box Attacks (Table \ref{table:adaptive}). We observe that if the attacker's NVM model is different from the target, the attacks do not transfer well and are weaker than non-adaptive attacks. For example, for CIFAR-10, under attack epsilon $\epsilon=1/255$, the accuracy of 64x64\_300k NVM model is $0.60\%$ for a non-adaptive attack, but $43.45\%$ for an adaptive attack with incorrect NVM model. Thus having an incorrect crossbar model is worse than having no crossbar model at all in this case.

\section{Discussion}

Non-idealities in NVM crossbars have been a long-standing challenge \cite{chakraborty2020geniex} which cause accuracy degradation in DNNs, and several techniques have been proposed to compensate for it \cite{chakraborty2018technology,cxdnn}. In this work, we study these non-idealities from the new perspective of adversarial robustness. We observed that DNNs implemented on an NVM crossbar hardware exhibit increased adversarial robustness under varied threat scenarios. While this robustness falls short of other defenses \cite{guo2017countering, dhillon2018stochastic, xie2017mitigating}, an important point to note is that such robustness is intrinsic to the NVM crossbar hardware, unlike other defenses which have a computational overhead. Also, any algorithmic defense can be further implemented on the analog hardware for additional robustness. The non-ideality factor (NF) of the crossbar model determines the degree of robustness. Therefore, one can potentially design NVM crossbars with optimal trade-off between accuracy degradation and increased robustness due to non-idealilties. We have demonstrated "Hardware-in-Loop" attacks where the knowledge of underlying hardware helps generate stronger attacks. While we have considered NVM crossbar models based on RRAM technology \cite{wong2012metal}, analog hardware based on other technologies \cite{wong2010phase, fong2015spin} are also possible. This, along with chip to chip variations, may further hinder the transferability of attacks generated on one analog computing hardware to another. In summary, this work is the first step toward understanding the role of non-idealities in NVM crossbar hardware for adversarial robustness. It opens the possibilities of defenses that leverage the non-ideal computations, and on the other hand, attacks that exploit these non-idealities. 


\bibliographystyle{IEEEtran}
\vspace{-3mm}
\bibliography{references}

\end{document}